\documentclass[aip,apl,reprint,preprintnumbers,amsmath,amssymb,floatfix,longbibliography, twocolumn]{revtex4-1}
\usepackage{graphicx}
\usepackage{dcolumn}
\usepackage{bm}
\usepackage{color}
\usepackage{xcolor}
\usepackage{tabularx}
\usepackage{amsmath,amsfonts,amsthm,amssymb}
\usepackage{appendix}
\usepackage[bookmarksnumbered,pdfpagelabels=true,plainpages=false,colorlinks=true,linkcolor=blue,citecolor=blue,urlcolor=blue]{hyperref}
\usepackage[english]{babel}

\begin{document}
\title{Mutual Synchronization of Spin-Torque Oscillators within a Ring Array }

\author{ M. A. Castro}

\affiliation{Universidad de Santiago de Chile (USACH) Departamento de F\'isica, CEDENNA, Avda. Ecuador 3493, Estaci\'on Central, Santiago, Chile.}
\affiliation{Univ. Grenoble Alpes, CEA, CNRS, Grenoble INP, SPINTEC, 38000 Grenoble, France.}

\author{ D. Mancilla-Almonacid}

\affiliation{Universidad de Santiago de Chile (USACH) Departamento de F\'isica, CEDENNA, Avda. Ecuador 3493, Estaci\'on Central, Santiago, Chile.}

\author{B. Dieny}

\affiliation{Univ. Grenoble Alpes, CEA, CNRS, Grenoble INP, SPINTEC, 38000 Grenoble, France.}

\author{S.Allende}

\affiliation{Universidad de Santiago de Chile (USACH) Departamento de F\'isica, CEDENNA, Avda. Ecuador 3493, Estaci\'on Central, Santiago, Chile.}

\author{L. D. Buda-Prejbeanu}
\author{U. Ebels}

\affiliation{Univ. Grenoble Alpes, CEA, CNRS, Grenoble INP, SPINTEC, 38000 Grenoble, France.}

\begin{abstract}
An array of spin torque nano-oscillators (STNOs), coupled by dipolar interaction and arranged on a ring, has been studied numerically and analytically. The phase patterns and locking ranges are extracted as a function of the number $N$, their separation, and the current density mismatch between selected subgroups of STNOs. If $N\geq 6$ for identical current densities through all STNOs, two degenerated modes are identified an in-phase mode (all STNOs have the same phase) and an out-of-phase mode (the phase makes a 2$\pi$ turn along the ring). When inducing a current density mismatch between two subgroups, additional phase shifts occur. The locking range (maximum current density mismatch) of the in-phase mode is larger than the one for the out-of-phase mode and depends on the number $N$ of STNOs on the ring as well as on the separation.  These results can be used for the development of magnetic devices that are based on STNO arrays.  
\end{abstract}

\maketitle

Spin torque nano-oscillators (STNOs) are nanoscale signal sources that can convert a DC input signal (current or voltage) into a microwave output voltage signal\cite{Slonczewski, Chen}. Depending on the magnetization configuration of the polarizing and the free layer, an STNO can generate \color{black} rf signals in the 100 MHz to several tens of GHz range.  An important property of STNOs is their strong coupling between the oscillation amplitude and phase \cite{SlavinIEEE} which enables the tuning of their frequency via the DC input signal\cite{Kiselev, Rippard1}. Furthermore, it enables, via an additional time-varying input signal, injection locking of the STNO frequency and phase\cite{Rippard2}, modulation of the STNO amplitude, frequency or phase \cite{Pufall, Litvinenko1}, or sweep-tuning of the STNO frequency \cite{Litvinenko2}. These properties open a large range of potential applications such as wireless communication \cite{Choi,Litvinenko1}, ultra-fast spectrum analysis \cite{Litvinenko2} as well as oscillator based hardware implementations for neuromorphic computation\cite{Grollier, Yogendra}. For these applications, but also from a fundamental point of view, the collective excitation states of a small or larger sized array of coupled STNOs is of interest\cite{Lebrun, Tsunegi}. Here, STNOs offer a rich variety of   coupling mechanisms (e.g. electrical\cite{ECurrent, Turtle, Chimera}, dipolar\cite{Lebrun,Tsunegi, Araujo, Flovik, DMancilla} or spin pumping\cite{Taniguchi2019}) and coupling scenarios (long range with all-to-all coupling or short range with nearest neighbors coupling). 

\begin{figure*}[htp]
    \centering
    \includegraphics[width=16cm]{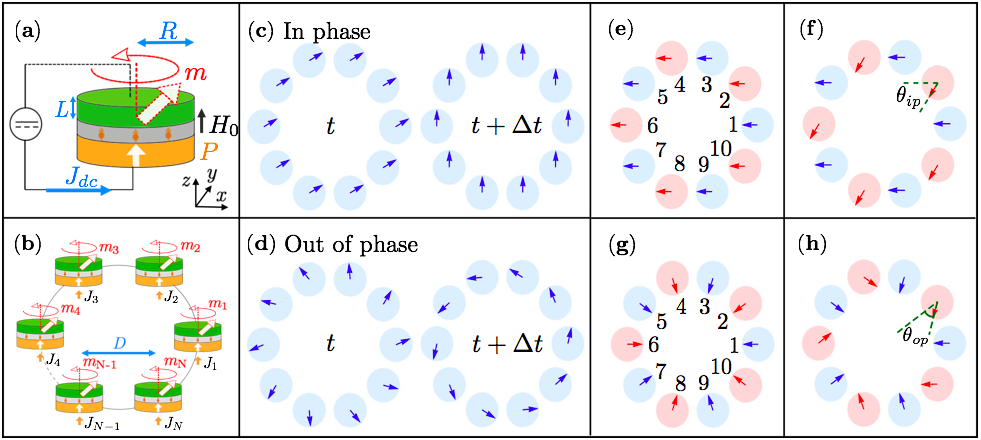}
    \caption{Schematic of (a) a single STNO and (b) an array of $N$ STNOs distributed along a ring. Phase states of the $N=10$ synchronized STNOs ring array for identical current densities through the STNOs: (c) in-phase mode and (d) out-of-phase mode. Phase patterns for non-identical current densities through the STNOs applied to: (e), (f) the in-phase mode and (g), (h) the out-of-phase mode. (e), (g) are for zero current density mismatch and show the labeling of STNOs. (f), (h) are for non-zero mismatch where the additional phase rotation is indicated by the angles $\theta_{ip}$ and $\theta_{op}$. The blue color illustrates the odd (1,3,5..) and the red color the even (2, 4, 6..) STNO position within the ring array. 
    }
    \label{newfig1}
\end{figure*}

A general question for such coupled STNO arrays is under what conditions a stationary fully coherent dynamic state exists, for which the phases of all STNOs are correlated and for which the corresponding phase differences are constant in time (as compared to the free running state where the phases are free and uncorrelated). Other solutions for the collective state might exist such as chaotic states \cite{ECurrent} or chimera states \cite{Chimera}, that will depend on the number of STNOs (large arrays), the coupling type, external control parameters (current, field) as well as on the homogeneity of the STNO properties (identical STNOs vs. a dispersion of STNO parameters). These states have been observed in other systems like 2D periodic lattice of Kuramoto oscillators \cite{Sarkar2021} or ring structure of identical Kuramoto oscillators \cite{Dnes2019}.
Previous simulation studies on two STNOs\cite{DMancilla}, coupled through dipolar interaction, have shown that depending on the initial condition or STNO configuration, it is  possible to stabilize an in-phase or anti-phase synchronised state \cite{Taniguchi2017,Wang2018,Taniguchi2018,Nakada2012}. Such modes can be associated with different applications, for instance, the in-phase mode is required to enhance the emission power, and the anti-phase mode can be useful for applications in phased array radar systems or bio-inspired computing \cite{Taniguchi2019,Kudo_2017}. Then, a general question is whether the number of possible phase states increases upon increasing the number $N$ of STNOs and how to control the phase patterns via frequency mismatches. Here we make a first step in this direction, and report on the different phase states that can be obtained for a small STNO array (number $N$=4-12) where the STNOs are arranged in a ring array, see Fig. \ref{newfig1}. Such a ring can be viewed as a 1D line with periodic boundary conditions. We consider as the interaction type all-to-all dipolar interaction of identical STNOs and determine the solutions for the phase states numerically as well as analytically. We report in-phase and out-of phase solutions obtained for different initial conditions when an identical current density is applied to all STNOs. Further phase states are obtained when the current density in a subgroup of STNOs is varied while it is kept constant in another subgroup.

For the modeling, STNO nanopillars of circular shape with radius $R$ and of free layer thickness $L$, (see Fig. \ref{newfig1}(a)) are considered that support an out-of-plane precession (OPP) mode. This is achieved by a perpendicular polarizer and a free layer that is in-plane magnetized. In this case, the magnetization $\bm M$ oscillates around the out-of-plane z-axis\cite{Houssameddine} providing very symmetric oscillation trajectories that can be modeled analytically. Furthermore a strong out-of plane field $\bm  H_0$ is applied to saturate the magnetization $\bm M$ out-of-plane in absence of current. In this way, the oscillation amplitude and with this the frequency and the dynamic dipolar interaction fields increase with increasing DC current density \cite{Zhu}. For the ring structure, an even number $N$ of STNOs ($N=4, 6, 8, 10, 12$) is considered, with a center to center distance $D$ between two adjacent oscillators, see Fig. \ref{newfig1}(b). Upon increasing the number $N$, the ring diameter is increased to keep $D$ constant.

The solutions for the phases of each STNO in the synchronized state were obtained by numerical simulation and are compared to analytical phase solutions. For the numerical simulations, the coupled Landau-Lifshitz-Gilbert-Slonczewski (LLGS) equations are solved, where the dynamics of the $k$-th free layer is described by
\begin{align}
    &\dfrac{d\bm{m}_k}{dt} = -\gamma_0 (\bm{m}_k\times \bm{H}_{eff}^{k})+\alpha (\bm{m}_k\times \dfrac{d\bm{m}_k}{dt})\nonumber\\
    &-\gamma_0 a_j J_k \bm{m}_k \times(\bm{m}_k\times \bm{P}),
    \label{LLGS}
\end{align}
where ${\bm m}_k={\bm  M}_k/M_s$ is the normalized magnetization of the $k$-th free layer and  ${M_s}$ is the saturation magnetization;  $\gamma_0=\gamma \mu_0$, $\gamma$ is  gyromagnetic ratio and $\mu_0$ is the vacuum permeability; $\alpha$ is the damping constant; $J_k$ is the current density on the $k$-th STNO; $a_j = \hbar \eta/( 2 e\mu_0 M_s L)$ is the spin-transfer torque parameter; ${\bm P}=\hat z$ is the spin polarization direction and $\eta$ the spin polarization of the current. The effective field  on the $k$-th free layer is  
\begin{align}
    \bm{H}_{eff}^{k}=H_0 \bm{\hat z} +\bm{H}_{d} + \bm{H}_{int}^{k},
\end{align}
where $H_0$ is the external out-of-plane field, $\bm{H}_{d} $ is the self-demagnetizing field and $\bm{H}_{int}^{k}$ is the magnetostatic interaction field between STNOs (see Appendix).

The solutions are found for the following device and material parameters: $R=50$nm, $L=3$nm, $D=160$nm, $M_s = 10^6 \text{A}/\text{m}$, $\alpha= 0.02$, $\mu_0 H_0 = 1.2 $T, \textbf{$a_j=3.78\times 10^{-8} $}m. The corresponding critical current density at which steady state oscillations for a single isolated STNO set in is $J_{c} = -0.427\times 10^{11} \text{A}/\text{m}^2$. In the following only values of current densities are considered for which the STNOs are in the steady state.

The first case of interest is that of identical current density applied to all STNOs. Two different phase patterns were identified by numerical LLGS integration as shown in Fig. \ref{newfig1}(c,d): (c) an in-phase mode with zero phase difference between adjacent STNOs $\triangle\phi^{ip}=\phi_{k+1}-\phi_{k}=0$ and (d) an out-of-phase mode characterized by a phase difference of $\triangle\phi^{op}=\phi_{k+1}-\phi_{k}=2\pi/N$. These two modes were obtained using as an initial condition a random distribution of the initial phases $\phi_{k}$ for each STNO. The in-phase mode is the only solution identified if the STNO number is $N < 6$. For larger arrays with $N\geq 6$, both the in-phase and out-of-phase modes are found, the in-phase mode being the most probable among them. Due to its symmetry, the out-of-phase mode is characterized by an overall zero dynamic in-plane magnetization, see Fig. \ref{newfig1}(d). Furthermore, despite the different phase patterns, the frequency and the total energy of the two modes are found to be equal within 1\% accuracy.

\begin{figure*}[htp]
    \centering
    \includegraphics[width=18cm]{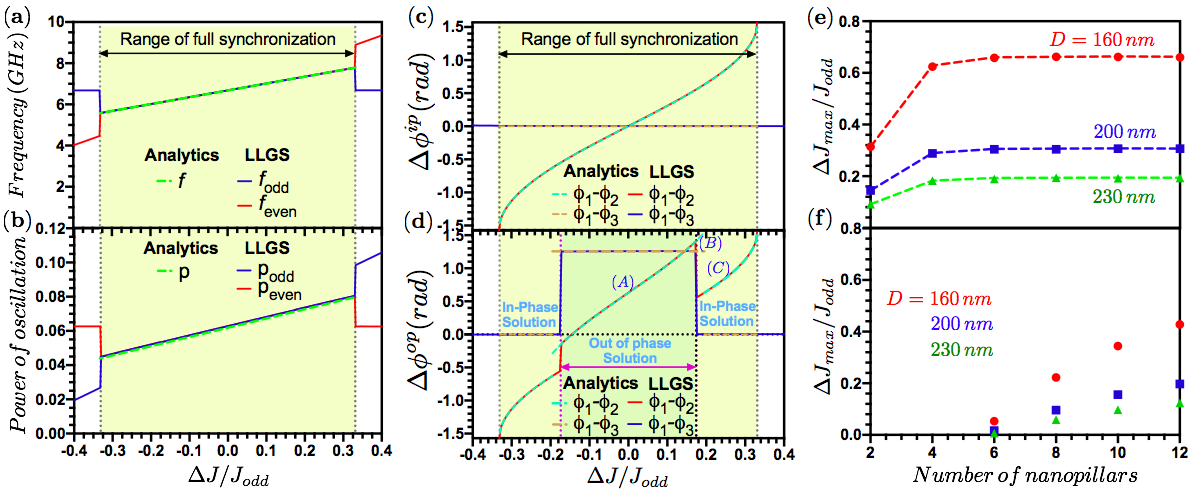}
    \caption{ Dependence of (a) the frequency and (b) the power of oscillation on the normalized current density mismatch $\Delta J$/$J_{odd}$ with $J_{odd}=-1 \times 10^{11} \text{A}/\text{m}^2$. In the macrospin simulations first the in-phase or out-of-phase modes were established at zero mismatch and then $J_{even}$ was varied. Both modes result in the same frequency and power vs. mismatch (up to 1\% accuracy). Phase difference $\Delta \phi $ as a function of the current density mismatch, when  starting the system (c) in the in-phase and (d) the out-of-phase state. The vertical gray dotted-lines indicate the limits of the full synchronisation range, where the phase difference between the closest STNOs is $\pm \pi/2$, see Eq.(\ref{eqInPhase}). The point (A) in (d) represents the starting-point of the macrospin simulations when the current density mismatch is zero. The point (B) in (d) corresponds to the transition between the out-of-phase to the in-phase solutions ($J_{even}=-1.174\times 10^{11}  \text{A}/\text{m}^{2}$). The point (C) in (d) corresponds to the situation when the phase difference between the STNOs follows the in-phase solution. Locking range in terms of maximum current density mismatch for (e) the in-phase mode and (f) the out-of-phase mode as a function of the number $N$ of STNOs and for different separations $D$ between STNOs. Full dots are results from the numerical simulations and dotted lines are from analytical solutions.}
    \label{newfig2}
\end{figure*}

In order to induce further phase patterns, the current densities were modified as follows (for both modes): two sub-groups of STNOs are formed, one with an odd label of STNOs ($1,3,5…, N-1$) within the ring for which $J_{odd}$ was kept constant, and one for even labels ($2,4,6,...,N$) of STNOs for which the current density $J_{even}$ has been varied (blue and red circles in Figs. \ref{newfig1}(e-h), respectively), increasing and decreasing its value with respect to the odd one $J_{odd}$, starting from the in-phase and out-of-phase mode pattern. Inducing a current density mismatch $ \Delta J = J_{even} - J_{odd}$ is equivalent to inducing a frequency mismatch $ \Delta f=f_{even} - f_{odd}$ between even and odd STNOs in the free running (uncoupled) state. As shown in Fig. \ref{newfig2}(a,b), it is observed that in the synchronized state (i) the frequency and the power are the same for all STNOs and for both modes; (ii) there exists an upper limit for the current density (frequency) mismatch $ \Delta J(\Delta f)$ that can be identified as the locking range for full synchronization, for which the two subgroups are synchronized together and have the same frequency, see Fig. \ref{newfig2}(a,b). For mismatches larger than the locking range, the even and odd subgroups are synchronized within each other (i.e., STNOs $1,3,...$,$N-1$ are synchronized and STNOs $2,4,...$,$N$ are synchronized), but the two subgroups are no more synchronized to each other and oscillate at different frequencies and have different power, see  Fig. \ref{newfig2}(a,b). This is an interesting result, meaning that it is possible to generate two independent subgroups of synchronized STNOs oscillating each one at their own frequency by imposing a specific pattern of current density distribution. Moreover, the synchronized STNOs of each subgroup are not nearest neighbors. Thus the subgroup is not a geometrical cluster, where the synchronized STNOs are located all in the same region of space, but their positions are rather intermixed. We expect that it is possible to extend this finding to more than two subgroups and to different spatial patterns of the current density mismatch. 

While inside the locking range  the frequency and power are the same for both modes, the phase patterns are different and evolve with increasing mismatch. As shown on Fig.\ref{newfig1} (f,h) the phase differences for the in-phase (out-of-phase) mode acquire an additional phase shift $\theta^{ip}$ ($\theta^{op}$) between odd (constant $J_{odd}$) and even (varying $J_{even}$) STNOs. In order to better quantify these additional phase shifts, analytical expressions were derived for the STNO phases for the two modes. These analytical expressions were obtained using the spin wave formalism\cite{SlavinIEEE}, that transforms the normalized magnetization vector of the free layer to a complex variable $c_k = (m_x^k+i m_y^k)/\sqrt{2(1+m_z^k)}$ and $c_k^{\dagger} = (m_x^k-i m_y^k)/\sqrt{2(1+m_z^k)}$.  This change of variable is usually defined through the Holstein-Primakoff transformation\cite{SlavinIEEE,DMancilla, Rezende}. It is convenient to write the complex-amplitude $c_k$ in terms of the power, $p_k$, and phase $\phi_k$ of oscillation, using  $c_k = \sqrt{p_k}e^{i\phi_k}$. By applying these definitions to the Eq. (\ref{LLGS}) and neglecting the non-resonant terms, it is possible to write $2N$ coupled equations for the power and phase of oscillation
\begin{align}
    &\dot{p}_k+ 2\Gamma_{eff}^{k}(p_k)p_k \nonumber\\
    &= -\sum_{l=1, k\neq l}^{N}\Omega_{k,l}\sqrt{p_k p_l} [2-(p_k+p_l)]\sin(\phi_k-\phi_l),\label{powerEq}
     \end{align}
    \begin{align}
    &\dot{\phi}_k+ \omega_k(p_k) \nonumber\\
    &= -\sum_{l=1, k\neq l}^{N} \dfrac{\Omega_{k,l}}{2} \sqrt{\dfrac{p_l}{p_k}}\left[2-(3p_k+p_l)\right]\cos(\phi_k-\phi_l).  \label{phaseEq}
\end{align}
Here $\Gamma_{eff}^{k}(p_k) = \Gamma_{+}^{k}(p_k)- \Gamma_{-}^{k}(p_k)$ is the non-linear effective damping, $\Gamma_{+}^{k}(p_k)$ and $\Gamma_{-}^{k}(p_k)$ are terms related to the dissipation of energy and the injection of energy induced by the spin current, respectively; $\omega_k(p_k)$ is the nonlinear frequency and $\Omega_{k,l}$ is the coupling constant that depends on the center-to-center distance $D_{k,l}$  between the STNOs $k$ and $l$ (see the Appendix for the definition of these parameters). In general, it is difficult to find an analytic solution to the system of Eqs. (\ref{powerEq}) and (\ref{phaseEq}) since they are strongly coupled. However, when the system is fully synchronized, the power of oscillations is practically constant and the same for each STNO as confirmed from the numerical simulations (see Fig.\ref{newfig2}(b), full lines).

Then, setting $p_k=p$ and $dp/dt=0$, it is possible to re-write the system of Eqs. (\ref{powerEq})-(\ref{phaseEq}) as follows
\begin{align}
    &\Gamma_{eff}^{k}(p) = -(1-p)\sum_{l=1, k\neq l}^{N}\Omega_{k,l} \sin(\phi_k-\phi_l),\label{powerEq2}\\
    &\dfrac{d\phi_k}{dt}+ \omega_k(p) =-(1-2p) \sum_{l=1, k\neq l}^{N} \Omega_{k,l} \cos(\phi_k-\phi_l). \label{phaseEq2}
\end{align}

When the STNOs are fully-synchronized, we obtain the condition $\sum_{k=1}^N\Gamma_{eff}^{k}(p) =0$ from the Eq. (\ref{powerEq2}), and then, the power of oscillations is given by 
\begin{align}
    p = -[a_j J_{av}+\alpha(H_0-H_d)]/(2\alpha H_d),
    \label{poderOscilacion}
\end{align}
where $J_{av}=(J_{odd}+J_{even})/2$. This is the same relation as for the power of a single oscillator, replacing the current density $J$ by the average $J_{av}$.

When the current density $J_{even}$ varies, the in-phase (ip) mode and out-of-phase (op) mode satisfy respectively:
\begin{align}
    &\phi_{k}^{ip}-\phi_{k+2}^{ip}=0, &&\phi_{k}^{ip}-\phi_{k+1}^{ip}=(-1)^{k}\theta^{ip},\label{faserel}\\
&\phi_{k}^{op}-\phi_{k+2}^{op}=\dfrac{4\pi}{N}, &&\phi_{k}^{op}-\phi_{k+1}^{op}=\dfrac{2\pi}{N}+(-1)^{k}\theta^{op}.\label{outrel}
\end{align}

By using Eqs. (\ref{faserel}) and (\ref{outrel}) in Eq. (\ref{powerEq2}), we obtain an expression for the additional phase shifts 
\begin{align}
    &\theta^{ip/op} = \arcsin\left(\dfrac{\gamma_0 a_j \Delta J}{2  \sum_{k=1}^{N/2} \Omega_{1,2k} g^{ip/op}(k)}\right), \;
    \label{eqInPhase}
\end{align}
where $g^{ip}(k)=1$ and $g^{op}(k)=2\pi\cos(2k-1)/N$. These solutions are Adler-type equations \cite{Pikovsky} and exist if $\vert \sin (\theta_{ip/op})\vert \leq 1$.

Eq. (\ref{phaseEq2}), also provides an expression for the frequencies of the STNOs for the  in-phase ($f^{ip}$)  and  out-of-phase ($f^{op}$) synchronization modes, where $2\pi f=d\phi/dt$
\begin{align}
&f^{ip/op} = -a_j \gamma_0 (J_{av}+\alpha^2 J_{even}) /(2  \pi\alpha)\\
&-\gamma_0\dfrac{(a_j  J_{av}+\alpha H_0)}{2\pi\alpha } \sum_{k=2}^{N} \dfrac{\Omega_{1,k}}{ \gamma_0 H_d}  \cos(\phi_1^{ip/op}-\phi_k^{ip/op})\nonumber\;.
\end{align}

The frequencies $f^{ip}$ and $f^{op}$ are practically identical. In the first term of the right side we can neglect the term proportional to $\alpha^2$. Additionally, the second term of the right side is related to the interaction and it is proportional to $\Omega_{1,k}/(\gamma_0 H_d)\sim 10^{-4}$ for next nearest neighbors ($k=2$) and for the parameters used here. This term can thus also be neglected. As a consequence the in-phase  and  out-of-phase  modes have the same frequency $f$, Eq. (\ref{frecuencia}), that is, similar to the power (Eq. (\ref{poderOscilacion})), the same expression as for a single oscillator replacing the current density by the average of the two current densities. With this the frequency is also independent of the relative phases of the dynamic magnetization between adjacent STNOs.
\begin{align}
    f \approx  -a_j \gamma_0 J_{av}/(2   \pi\alpha ).\;
    \label{frecuencia}
\end{align}
This analytical derivation is in good agreement with numerical simulation results, see Fig. \ref{newfig2}(a, b), where the frequency and power of oscillations are shown to vary linearly with the current density mismatch. 

In the following, the analytical expressions for the phases Eq. (\ref{eqInPhase}) are analyzed in more detail and are compared to the results of the numerical simulations for the example of N=10 and a constant current density $J_{odd}=-1 \times 10^{11} \text{A}/\text{m}^2$ for odd STNOs. Figure \ref{newfig2}(c,d) shows the phase differences between the odd and even oscillator subgroups as a function of the normalized current density mismatch obtained from the LLGS numerical solution (full lines) and the analytical solutions (dashed lines).  Figures \ref{newfig2}(c) and \ref{newfig2}(d) show the phase differences when the system evolves respectively from the zero-mismatch ($\Delta J$ =0) in-phase or out-of-phase mode. As can be seen from Fig. \ref{newfig2}(c) in the case of the in-phase mode, the phase difference between all odd STNOs (1,3,5) is zero, while between the even and odd subgroups it is zero only for zero mismatch. For non-zero mismatch the phase differences follow an arcsine behaviour, Eq. (\ref{eqInPhase}) within the locking range (the green region in Fig. \ref{newfig2}(c)). At the locking boundary where the synchronization between the two subgroups is lost, the phase difference is $\pm \pi/2$. Beyond this range, the system is partially synchronized. The results for the phase difference from the analytical and the numerical simulations are in good agreement, and therefore, we demonstrated the validity of the analytical expression. 

When the system starts in the out-of-phase state, see point A in Fig. \ref{newfig2}d, we can distinguish two locking ranges. First, for zero mismatch, the phase difference between neighboring even and odd STNOs is non-zero and takes the value of ${2\pi/N=0.63 rad}$ (Eq. (\ref{outrel}). Increasing the mismatch, the phase of the even STNOs acquires an additional phase shift as given by Eq. \ref{eqInPhase}) with an arcsine dependence as a function of mismatch. In this range the results from numerical simulations and the analytical expressions agree well. The simulations show, that at a certain critical value of the current density mismatch (see point B in Fig. \ref{newfig2}(d), the out-of-phase mode becomes unstable and transits irreversibly into the in-phase mode characterized by the phase differences defined by Eq. (\ref{faserel}) (see point C in Fig. \ref{newfig2}d), until the phase difference between the even and odd STNO subgroups reaches $\pm \pi/2$ and the two subgroups are no more synchronized. Decreasing then the current density mismatch, the system remains in the in-phase mode and it is not possible to return to the out-of-phase state. It is noted, that the analytical solutions lead to a somewhat larger value for the critical current density mismatch where the out-of-phase mode transits to the in-phase mode. 

From the analytical expressions Eq. (\ref{eqInPhase}), one can derive an expression for the locking range of the in-phase mode. Synchronization is lost when the phase difference between the even and odd subgroups becomes $\theta_{ip}=\pi/2$. This leads to:
\begin{align}
    (J_{max}-J_{min}) = \dfrac{4}{a_j \gamma_0} \sum_{k=1}^{N/2}\Omega_{1,2k}.
    \label{locking}
\end{align}
Similar to injection locking of an oscillator to an external signal, the locking range here depends strongly on the coupling constant $\Omega_{k,l}$. It increases with the coupling, and thus Eq. (\ref{locking}) is the equivalent to Arnold tongue boundaries. Furthermore, Eq.(\ref{locking}) shows that the locking depends on the sum over dipolar interactions, and it is thus expected to increase with increasing number $N$ of STNOs in the ring. This is confirmed in Figure \ref{newfig2}(e), where the locking range is shown vs. $N$ from numerical and analytical calculations. Both agree well and show first an increase up to $N=6$ and then a saturation. This demonstrates that, for the range of values of $D$ considered here, the interaction is not limited to nearest neighbors, but that a larger range of STNOs needs to be considered.  Finally, since the dipolar interaction increases with decreasing separation $D$ between STNOs, the locking range strongly increases upon reducing $D$, as can be seen in Fig. \ref{newfig2}(e). From the numerical simulations, we have also extracted the current density mismatch where the out-of-phase mode transits to the in-phase mode. The corresponding locking ranges show a strong increase with the number $N$ of STNOs, see Fig.\ref{newfig2}(f) and a strong increase for decreasing separation.\color{black} 

 To conclude, we have studied theoretically the phase patterns that can be obtained for an array of STNOs arranged on a ring array and coupled via dipolar interaction. When the same current density is applied to all STNOs, two different synchronized modes are observed, the in-phase mode for which all STNOs have the same phase, and the out-of-phase mode where the phase makes a $2\pi$ turn along the ring array. The latter is observed only when the number $N$ of STNOs within the ring is larger or equal to $N=6$. Further phase patterns were obtained when varying the current density in a subgroup of STNOs, where all even STNOs formed the subgroup. For the in-phase mode, a current density mismatch (equivalent to a frequency mismatch), the two even and odd STNO subgroups remain synchronized within a certain synchronization range whose size scales with the dipolar interactions. Beyond this locking range, the two sub-groups de-synchronize, but the STNOs remain synchronized within each subgroup. The existence range of the out-of-phase mode is smaller than the in-phase synchronization range. Thus, a critical current density mismatch exists, where the out-of-phase mode becomes unstable and transits to the in-phase mode. Finally, despite the different phase patterns of the in-phase and out-of-phase mode (that lead to different dynamic dipolar interactions between adjacent STNOs), the frequency and power of the STNOs are the same inside the full synchronization range. The numerical results are confirmed from analytical results obtained by solving the phase equations of the coupled STNO array. Expressions have been provided for the ring array for the phase equation, the phase shifts as a function of current density mismatch, the power, frequency and Arnold tongue boundaries.
It is expected that upon increasing the number $N$ of STNOs within the array for identical current densities, it should be possible to induce further phase patterns, where the phase makes more than one turn along the ring. Furthermore, it is expected that a large variety of different phase patterns can be induced by choosing different subgroups of STNOs, for which the current density is varied. These results will be of interest for oscillator based computing applications. 

\section{Appendix}

The magnetic dipolar interacting field over the $k$ free layer of the STNO due the others  free-layers  is \cite{DMancilla}
\begin{align}
    &\bm{H}_{int}^{k} = M_s \sum_{l=1,l\neq k}^{N}  \Big\{\left([K_1+K_2\cos(2\varphi_{k,l})] m_l^{x}+\right.\nonumber\\ &\left.K_2\sin(2\varphi_{k,l}) m_l^y\right) \mathbf{\hat{x}}
    + \left([K_1-K_2\cos(2\varphi_{k,l})] m_l^y +\right.\nonumber\\
    &\left.K_2\sin(2\varphi_{k,l}) m_l^x\right)\mathbf{\hat{y}} +  K_3 m_l^{z}\mathbf{\hat{z}}\Big\}
\end{align}
where $\varphi_{k,l}$ is the angle between the STNOs $k$ and $l$, and  $K_i$ are functions that depend on the radius of the STNOs and on the inverse of the separation between them, see Eqs. B14 in Ref. \cite{DMancilla} for expressions of $K_i$. The coupling constant $\Omega_{k,l}$ depends on $K_1$ and is defined as:
\begin{align}
 &\Omega_{k,l} = \dfrac{M_s \gamma_0}{\sqrt{1+\alpha^2}} K_1(D_{k,l}) \approx M_s \gamma_0 K_1(D_{k,l})\
\end{align}

The non-linear parameters of Eqs. (\ref{powerEq}),(\ref{phaseEq})  are defined as follows: 


\begin{align}
   & \omega^k(p)=\omega^k_0+p^k {\cal{N}}\nonumber\\
   &\omega^{k}_{0} = \dfrac{\gamma_0}{1+\alpha^2}( H_{d}-H_0-M_s \sum_{j=1, l\neq k}^{N} K_3^{k,l} )\nonumber \\
   &{\cal{N}} = -\dfrac{2 \gamma_0}{1+\alpha^2} H_d 
\end{align}

\begin{align}
  &\Gamma_{+}^k (p_k) = - \dfrac{\gamma_0 \alpha}{1+\alpha^2} ( H_d-H_0-M_s\sum_{l=1, j\neq k}^{N} K_3^{k,l} )\nonumber\\
  &+\dfrac{\gamma_0 \alpha}{1+\alpha^2}  ((-H_0+3 H_d) p_k- H_d p_k^2)  \nonumber \\
  &\Gamma_{-}^k (p_k) = \dfrac{\gamma_0 a_j J_k}{1+\alpha^2} (1-p_k)
  \label{Gamma}
\end{align}

The power of oscillations, when the STNOs are synchronized, can be obtained by solving the equation $\sum_{k=1}^{N}\Gamma_{eff}^{k}(p)=0$, replacing the equation \ref{Gamma}, we find:

\begin{align}
p &= -\dfrac{aj J_{ave}+\alpha(H_0-3 H_d)}{4 H_d \alpha}\nonumber\\
&-\dfrac{\sqrt{(a_j J_{ave}+\alpha(H_d+H_0))^2+8 H_d \alpha^2 M_s\sum_{j=1, l\neq k}^{N} K_3^{k,l}}}{4 H_d \alpha}
\label{power_full}
\end{align}

where $H_d =  M_s (N_z-N_x)$. $N_z$ and $N_x$ are the corresponding demagnetization factors of the circular free layer. In general, the term $H_{d}-H_0 >>K_3$, in that way, it is possible to neglect the contribution of $K_3$ in the equation \ref{power_full}. This assumption only leads to a small shift between the numerical and analytical solution shown in Fig \ref{newfig2}.(b).   We would like to point out, that different to single STNO equations, the term proportional to $p_k^2$ cannot be neglected in Eq. \ref{Gamma}, in order to describe the synchronized state correctly.

\section*{ACKNOWLEDGMENTS} 

We acknowledge financial support in Chile from FONDECYT
Grants No. 1200867 and 1190727,
and Financiamiento Basal para Centros Científicos y Tecnológicos
de Excelencia FB 0807 (AFB 180001).  D.M.-A. acknowledges
Postdoctorado FONDECYT  3180416 2018 and Proyecto Postdoc DICYT, Código 042131AP POSTDOC, Vicerrectoría de Investigación, Desarrollo e Innovación.
M.A. Castro acknowledges Conicyt-PCHA/Doctorado Nacional/2017-21171016. This work was supported in part by the ERC Grant MAGICAL (N\textsuperscript{o}669204).

\end{document}